# The Design of Life:
# Information Technology, "Knapsack" and "Synprops" Used to Engineer the Sequence of Amino Acid Residues in Proteins

S. Bumble, Dept. of Physics, Community College of Physics, stan2@mailexcite.com

*"What sort of molecular code could be so elaborate as to convey all the multitudinous wonder of the living world?"* J. D. Watson, DNA, Alfred A. Knopf, New York, 2003

*"Today we are learning the language in which God created life. With this profound new knowledge, humankind is on the verge of gaining immense, new power to heal."* Bill Clinton, June 26, 2000

*"What is Life?"* Erwin Schrodinger

**Abstract**

An In Silico model to relate the properties of proteins to the structure, sequence, function and evolutionary history of proteins is shown. The derived ideal sequences for amino acid residues in proteins can then be considered as attractors for structures of actual protein families and their functions. Hopfield networks[1] can then act as GCMs (General Content Addressable Memories) when they are trained to recall unique pre-determined states when presented with information associated with that state. The tertiary structure of proteins is determined by structurally enforced interactions between residues and they help to elucidate common motifs exhibited in proteins. The acceptance (or rejection) of substitution on the protein backbone conforms to a nonlinear response and can lead to emergent behavior.

### I. Introduction

The creation of life and its script can be described as a chemical plant, or an auto manufacturing assembly line with control rooms, flow sheets, networks, quality control and product lifetime expectations. There are almost $9 \times 10^9$ products walking around on the planet earth today. In these analogies, nothing in the present conceptions of molecular biology is denied. However the above analogies can help to lead to better solutions to problems in molecular biology, the understanding causes of disease, the amelioration of health, the process of drug discovery and the quality of life for the organisms who will be alive today and in the future. There is a need for the diversity of life and so the great combinatorial constitution of each of us is immense and this may lead to imperfections in each of us affecting our health and length of life expectations. However the "Plant Manager" and "Quality Control Manager" may give each of us the chance to enter upon life at our birth just as the products of the GMAC plant of General Motors allows many of their products to leave the plant, even with imperfect quality control scores. Yet this paper will try to ameliorate the health, treat the disease and ameliorate the quality of life for the organisms entering upon and existing in the state of life.

## II. Molecular Biology

DNA is a double helix with the nucleotides A, T, C and G permutated along each strand with A bonded to T and C bonded to G by hydrogen bonds. It translates and replicates to various proteins which have twenty different amino acids permutated along a –NH-CHR-C(O)- backbone (denoted as structure A) where R represents the various amino acid residues. The amino acids are constructed according to a triplet code from the DNA constitution of nucleotides. However the sequence placement of nucleotides along the protein skeleton is discussed in part VI (Evooptimization and Protein Sequence).

"In animal cells, every gene can make as many as 5 to 10 proteins, and every protein can have a different function. We need to understand the layer or filter between the genome and the proteome and the interactions of a large number of components, including protein, DNA and metabolites. Three categories of models must be built: models for metabolism, DNA regulation and cell signaling. Metabolism leads the way, regulatory networks are a couple of years behind that, and signaling is still very complete. To build the model of all the physical and chemical functions in the cell requires a new type of modeling approach; one that is based on network theory." We must also understand that what takes place in the cell is a matter of "structured chaos" and that the "biological mesoscale has its own structures, logic, rules and regulatory mechanisms."

"Nearly every biological process is catalyzed by a set of ten or more spatially positioned interacting proteins that undergo highly ordered movements in a machine-like assembly."[2]

## III. Properties of Amino Acids and Proteins

An amino acid has the structure $NH_2CHRCOOH$ where the carbon with the R group attached is termed the alpha carbon. The R group has twenty different variations. The protein is a polypeptide polymer which is denoted in section II above as structure A. The repeating unit of the protein molecules has such properties as volume, dipole moment, log partition coefficient, activity coefficient, hydrophobicity, solubility parameter, chi, molar refractive index, etc. These properties confer properties to the protein molecule as a whole. However, the protein molecule also has biological properties which become active when the protein folds into a native globular molecule.

The properties of the amino acid residues are shown below for each of the twenty amino acid residues as derived from the program described below as SYNPROPS.

| residue | dipole mo | neglogparc | V/20 | hydrophob | logactcoef | sol param | CHI | mol refract |
|---|---|---|---|---|---|---|---|---|
| ala | 1.63945 | 2.89983 | 3.74845 | -2.60426 | 6.577995 | 8.27661 | 35.3748 | 17.11992 |
| asn | 4.67705 | 5.21377 | 4.43545 | -1.44596 | 14.93858 | 16.16167 | 13.8594 | 25.02192 |
| asp | 3.07055 | 4.40374 | 4.27155 | -7.37356 | 12.51525 | 13.07255 | 21.6879 | 23.64277 |
| cys | 2.38845 | 3.32977 | 4.2553 | -4.04259 | 9.367475 | 10.638 | 30.5513 | 23.50322 |
| gln | 4.56335 | 4.69192 | 5.2949 | -0.94134 | 14.83021 | 15.962 | 16.3845 | 29.53745 |
| glu | 2.95685 | 3.88189 | 5.131 | -6.86894 | 12.40688 | 12.872 | 24.2147 | 28.1583 |
| gly | 1.77355 | 3.42677 | 2.88915 | -2.93481 | 6.678975 | 8.48389 | 32.8367 | 12.61641 |
| iso | 1.34775 | 1.46208 | 6.456 | -0.4382 | 6.200525 | 7.39963 | 42.7831 | 30.72551 |
| leu | 1.34775 | 1.46208 | 6.456 | -0.4382 | 6.200525 | 7.39963 | 42.7831 | 30.72551 |
| lys | 2.06725 | 3.24955 | 6.4425 | -7.57275 | 10.6585 | 9.65799 | 32.0947 | 34.65878 |
| met | 2.31305 | 2.62797 | 6.00895 | -1.53475 | 8.950935 | 9.48157 | 35.9034 | 33.5721 |
| phe | 1.77705 | 1.47151 | 7.0071 | -3.35187 | 8.556295 | 9.13923 | 41.7417 | 41.19408 |
| ser | 2.88225 | 4.72784 | 3.4832 | -7.33078 | 11.63029 | 12.25083 | 21.2775 | 17.85461 |
| thr | 2.81825 | 4.333789 | 4.47185 | -6.17395 | 11.46956 | 11.77301 | 23.6356 | 22.42914 |
| val | 1.46145 | 1.98392 | 5.59655 | -0.94282 | 6.308895 | 7.59921 | 40.258 | 26.20998 |

Properties for the amino acid residues, as derived from SYNPROPS[3,4], were plotted in graphs shown below. The properties of the dipole moments (dipole mom.), log of the partition coefficients (log par. coef.) and the volumes (V/20) were the ones selected for the first graph. You will notice that the graph for the log of the partition coefficient qualitatively parallels that of the dipole moment and also that there is a block of amino acid residues whose volumes are above the other amino acid residues and above the values of the properties (dipole moments and log partition coefficients) chosen. Since no two objects can occupy the same space at the same time, this will be very important to avoid steric hindrance. These amino acids are iso, leu, lys, met and phe.

A glance at the parent structures of these amino acids in any elementary molecular biology text will corroborate their larger volume.

**Amino Acids**

| | | | |
|---|---|---|---|
| Glycine | Gly | G | H |
| Alanine | Ala | A | Me |
| Valine | Val | V | (CH3)2CH |
| Leucine | Leu | L | (CH3)2CHCH2 |
| Isoleucine | Ile | I | CH3CH2CH(CH3) |
| Phenylalanine | Phe | F | PhCH2 |
| Tyrosine | Tyr | Y | HOC6H4CH2 |
| Tryptophan | Trp | W | Indole-CH2 |
| Serine | Ser | S | HOCH2 |
| Threonine | Thr | T | CH3CH(OH) |
| Methionine | Met | M | CH3SCH2CH2 |
| Cysteine | Cys | C | HSCH2 |
| Asparagine | Asn | N | H2NCOCH2 |
| Glutamine | Gln | Q | H2NCOCH2CH2 |
| Proline | Pro | P | -CH2CH2CH2 |
| Aspartic acid | Asp | D | HOOCCH2 |
| Glutamic acid | Glu | E | HOOCCH2CH2 |
| Lysine | Lys | K | H2NCH2CH2CH2CH2 |
| Arginine | Arg | R | H2NC(=NH)NHCH2CH2CH2 |
| Histidine | His | H | Imidazole-CH2 |

In *Figure* 2 the solubility parameter (sol. param.) and the log of the activity coefficient (log act. coef.) have a good measure of parallel homology or correlation as also does the negative values of the log of the partition coefficient and the values of the dipole moment. The negative values of the hydrophobicity (hydrophob., or sometimes the hydrophilicity (hydrophil.)) is used, which is the negative of the hydrophobicity (negative of the hydrophobicity), however, stands alone. This leads to the observation that there are four classes of physical property variables for the amino acid residues they are: I. log of the partition coefficient and the dipole moment; II. Solubility parameter and the log of the activity coefficient; III. the molar refractivity and the chi factor and IV. The hydrophobicity or the hydrophilicity. This can be instrumental in determining what variable to use in QSARs. It is interesting to note also that there are four nucleotides in DNA that is the carrier of genes.

It was found that some times the correlations between physical properties for the amino acid residues were difficult to ascertain from the data. To achieve this objective, each physical property of the residues was sorted and a number assigned to the each property of the amino acid residues that was the ranking of the each property for each of the amino acid residues after the sort. This number or ranking for each of the amino acids for each of the properties was then substituted for the magnitude of the value of the property and this was used in each of the four graphs plotted below. It was also noted that the physical properties, dipole moment and the negative of the log of the partition coefficient (group A) did follow the same trends on the graphs after these plots were made of the ranking versus the residue. The same was true for the ranking of the log of the activity coefficient and the solubility parameter (group B) plotted versus the residue. Thus, one of the properties of the properties from those of group A and group B was used to represent all the properties shown in both groups A and B (physical property representation). This was justified as the original properties were never meant to represent the absolute magnitude of each of the values, but only to represent the relations of the magnitudes to each other within each property. Then we can group all the properties into only three classes: I). Property representation, II). V/20 and III). Hydrophilicity. This then is the formulation of the graphs shown in *Figures* 7 through 10 and led to observations such as that pronounced in *Figure* 10, where the plot is made according to the sort of the hydrophilicity and their residue assignment and the physical properties and the V/20 are plotted according to the residue assignment of the hydrophilicity-residue sort. It shows that the graphs of the V/20 and physical properties are very much in parallel or show the same trends for the first six residues: gly, ser, ala, cys, asp and thr. This then leads to the contention that the first evolutionary residues used to form the first proteins were these. It is also important to note that all these primitive residues contain but two carbons, not more as the other residues possess, further supporting this contention. It is also now tempting to consider a QSAR of the form $P = aP_I + bP_{II} + cP_{III}$, where $P$ represents the overall physical property $P_I$, $P_{II}$ and $P_{III}$ represent the physical property, the V/20 and the hydrophilicity, respectively.

Functional features of the parent conformation can be memorized or inherited and then manifested in other conditions. Functional properties are intertwined with molecular evolution in the earliest stages. Surfaces where polymers are adsorbed or contacted can imprint and mold by catalysis the sequence design of the ligand molecule. In this case the parent molecule can act as a template for the attached molecule. This can lead to functional properties of the ligand molecule after it is desorbed or separated. Then the features of the parent molecule are manifested in other conditions. This is a mechanism of molecular evolution. The polymer will acquire special primary sequences in the parent condition and extend this to other conditions using the fact that the primary structure is tuned to perform certain functions.

## IV. SYNPROPS

SYNPROPS uses the data and formulas of Cramer that are inserted into a spread sheet program such that once the composition of molecules is given in terms of molecular groups, the properties of designated molecules can be ascertained to a given level of accuracy. Cramer derived the data originally for the molecules in the handbooks of chemistry and physics and the predictions of properties of other molecules were deemed to be accurate and the statistics used in the whole process were above suspicion. The SYNPROPS matrix of data and formulas were also used in the usual optimizing programs contained in the PC spreadsheet programs of Quattro Pro, Excel, etc., so that one could find the best structures of molecules (and mixtures of molecules) for desired properties, and furthermore, the whole procedure was reduced to one where methods of matrix algebra could be utilized to proceed backwards and forwards in the manipulation of properties and structures of molecules. These methods were used for determining properties for biological molecules, such as amino acids and proteins and for applications such as in environmental substitutions, toxicity and drug discovery.

## V. KNAPSACK[5]

KNAPSACK is a program adopted from discrete mathematics and also is processed on a spreadsheet such as Excel or Quattro Pro. It calculates the optimum distribution of objects that can be contained in a "property container" if the property values for each object can vary to a different degree. Originally it calculated the frequency of each amino acid in a protein if the protein was limited to a maximum number of amino acids. Then it was also extended to also determine the practical appropriate sequence of these amino acids in the protein. The number of amino acids considered was fifteen instead of twenty because the data to regress the properties of the five remaining amino acids with their structure was not available at the time but this can be addressed in the future.

The KNAPSACK program utilizes the variables $n_i$ for the number of the amino acid residue $i$ and $P_i$ for the particular property for the amino acid residue $i$. Then the summation Sum ($i = 1$ to $i = 15$) for the function $S = n_i P_i$ from $i = 1$ to $i = 15$ is to be maximized or minimized according to what particular property $P$ represents. Now the summation of $N = n_i$ (from $i = 1$ to $i = 15$) is fixed according to the length of the peptide or the protein chain. This means the extreme of the of sum of $S = n_i P_i$ is found under the constraint that the sum of $n_i$ ($N$) is a fixed number. When this is done and the values of $n_i$ are allowed to vary while the extremum of $S = n_i P_i$ is found under the constraint that the sum of $n_i$ is a fixed value then the spectrum of the values of each of the variables $n_i$ found (where each of these values is the number of each particular amino acid residue), then this set of numbers is the frequency spectrum of these amino acids under the conditions that were set.

## VI. "Evooptimization" and Protein Sequence

Now although the frequency has been determined, we would also like to determine the sequence of the amino acid residues along the polypeptide backbone or the protein. We can also use the Knapsack program to do this if we infer that the sequence is determined by the interaction between each residue with its neighbors. This interaction can be determined by the relative properties of these neighbors and also the evolution of the sequence from its start. We call the process for doing so "evooptimization". This we start with any one of the fifteen amino acid residues attached to the alpha carbon atom. We fix the identity of this particular residue on the designated carbon atom. Then we restrict the total number of residues to the number two. We then find the extremum value for $S$ under the constraint that the total number of residues be $N$. However, there is another constraint; that of steric hindrance. Two objects may not occupy the same space at the same time, thus some factor of the molecular volume of the residue molecular volume that is the neighbor must be another constraint. By trial and error this factor is found to be $a$. Thus, if we choose either the property to be the dipole moment or the log of the partition coefficient we have the value of $n_j$ (the neighboring residue) constrained by its molecular volume to be a minimum while its effect at the same time is to maximize the dipole moment of the molecules or to minimize the log of the partition coefficient of the two residues. Thus we have $F_i = P_I - aV_j$ if $P$ is the dipole moment and $F_i = P_I + aV_j$ if $P$ is the log of the partition function. The difference in the sign occurs because dipole moments are positive in the required sense and log of the partition coefficients are negative. Thus the first equation above is to be maximized and the second equation is to be minimized.

We can then derive an equation for $F_I = a_1 P_I + a_2 P_{II} + a_3 P_{III} + a_4 P_{IV} - a_5 V$ which will take into account the four classes of physical properties found and the constraint of the steric hindrance of the volume.

Now when we add a third member to the sequence, $N$ becomes fixed at three, etc., until we find the whole sequence. This has been done for other properties as well, including hydrophobicity, log of the activity coefficient, solubility parameter, CHI, and molar refraction; all these property values derived from SYNPROPS as discussed above.

It now remains to find homologs of these sequences in the experimental sequences found in actual known proteins. Furthermore, the procedure can be carried out for DNA itself as this is a sequence of only four nucleotides C, A, G and T. This result is then to be compared with the DNA of actual organisms. To do this we need to know the properties of C, A, G and T. This is yet to be done by a SYNPROPS type of program if we can obtain the necessary data or calculate such data accurately. If we can do this then it is possible that we can engineer, in silico, genes, since they are an assembly of nucleotides

The procedure above can be seen as similar to a game of chess. A player does not think of the move of the chess pieces at hand but what it would mean to the board several moves in the future and perhaps to the final move. In the same way the evolution of the best fit for the sequence in a protein is anticipated as best heading; towards the final sequence that will fold the protein optimally to its most active function. This is planted in the "memory" by training the network to that consistent to a Hopfield network.

Now frequently the biopolymer, whether DNA or a protein, may be in a helix, coiled or globular form. In these cases an interaction of concern may not be between a link and that of its nearest neighbor but with a neighbor further removed. This happens frequently through hydrogen bonds. In such cases, KNAPSACK can also function efficiently. In such cases, KNAPSACK uses its ability to impose constraints in its maximizing ability such that the interactions can be examined for not only nearest neighbors but also with neighbors that are *n* links removed when such calculations are carried out in parallel for all configurations.

### VII. Homology with Actual Proteins and Their Motifs

"Looking at the similarity between the human genome and other species is a really powerful way to get at functional sequences and allow us to work on them in different species."[6] Sequence similarity searching is used frequently by biologists. BLAST (Basic Local Alignment Search Tool) is most popular for the comparisons between pairs of sequences and searches for regions of local similarity. It is based on dynamic programming methods first applied in biological work by Needleman and Wunsch. The Smith-Waterman algorithm is an exhaustive and mathematically optimal method. Yet it is too slow for exhaustive searches for huge nucleotide sequence repositories such as GenBank or amino acid databases like SWISS-PROT. Therefore heuristics are employed to approximate the best local alignment which they often do. Now that the human genome is 'complete', the sequence, structure and quantity of each protein present is necessary to complement the genomic and transcriptional data. Phylogenetic profiling or assigning to a protein a string that identifies its presence or absence in every known genomic sequence, on the basis of the observation that functionally linked proteins are either preserved or eliminated in a new species is called correlated evolution. Data mining is also used. It is a search for hidden trends within large sets of data.

On comparing primate proteins we find amino acid sequence in the hemoglobin of primates;

| | | | | | | | | | |
|---|---|---|---|---|---|---|---|---|---|
| Human Being | SER | THR | ALA | GLY | ASP | GLU | VAL | GLU | ASP | THR |
| Chimpanzee | SER | THR | ALA | GLY | ASP | GLU | VAL | GLU | ASP | THR |
| Gorilla | SER | THR | ALA | GLY | ASP | GLU | VAL | GLU | ASP | THR |
| Baboon | ASN | THR | THR | GLY | ASP | GLU | VAL | ASP | ASP | SER |
| Lemur | ALA | THR | SER | GLY | GLU | LYS | VAL | GLU | ASP | SER |

While for some Non-Primates

| | | | | | | | | | |
|---|---|---|---|---|---|---|---|---|---|
| Dog | SER | SER | GLY | GLY | ASP | GLU | ILE | ASP | ASP | SER |
| Chicken | GLN | THR | GLY | GLY | ALA | GLU | ILE | ALA | ASN | SER |
| Frog | ASP | SER | GLY | GLY | LYS | HIS | VAL | THR | ASN | SER |

Also pertinent is the work of Christian de Duve[7] who examines the amino acid sequence in the protein enzyme phosphoglycerate kinase (PKG) for similarities in the colibacillibus, wheat, the fruitfly, the horse and human. These can be analyzed with phylogenetic tree building and comparative sequencing to show that the human/fruit fly antedates the human/horse bifurcation by a time factor of 6.

Watson quotes in his book, DNA, "the more closely related two species are in evolutionary terms, the more similar are the sequences of their corresponding proteins. It is noted that, in one of the protein chains of hemoglobin molecules, over its total length of 141 amino acids, there is only one difference between the human version and the chimpanzee, but the difference between humans and horses is 18 amino acids."

Now if we study the physical properties of each amino acid residue in the sequence of the human being, we find that when each of the properties are ranked the number of the ranking for four of the physical properties (dipole moment, log of the partition coefficient, the log of the activity coefficient and the solubility parameter) is the same. Furthermore, as we proceed up the evolutionary scale, in general, the hydrophilicity goes down in rank, whereas the volume goes up. It is as if the proteins, as they develop, squeeze out the water from the native form and learn that they can accommodate higher volumes of the amino acid residues in their sequence.

## VIII. Biological Networks

Biological networks can be constructed with a given degree (the number of edges incident to a vertex), the minimum path distance between pairs of vertices and the clustering coefficient (the fraction of edges among the neighbors of a vertex). Such topological characteristics may proceed from evolutionary rules that contain preferential attachment. However, local models of growing networks can also explain strong similarities in topological properties.

Three different network applications are being studied in this work. They are metabolic networks, protein interaction networks and genetic regulatory networks. The metabolic networks represent "metabolic substrates and products with directed edges joining them. If a known metabolic reaction exists that acts on a given substrate and produces a given product."[8] In this regard, the work in metabolic engineering is cited and the powerful method of Fan et alia[9] is to be regarded as essential in emphasizing the elucidation of the fundamental reactions and cycle pathways.

Also of significance is the network of mechanistic physical interactions between proteins (as opposed to chemical reactions among metabolites), which is usually referred to as protein interaction networks.

An interesting article by Wingreen et alia[10] links designability (of folds having lowest energy of an unusually large number of sequences) of proteins with thermal stability and attributes this to the dominance of the hydrophobic solvation energy.

The third kind of network is the class of biological networks known as genetic regulatory networks. They were the first networked dynamical systems for which modeling attempts were made using random Boolean nets by Kauffman.[11] Genetic regulatory networks now include the expression of genes, a production (the transcription and translation of the protein for which the gene codes) that can be controlled by the presence of proteins, both activators and inhibitors, so that the genome itself forms a switching network with vertices representing the proteins and directed edges representing the dependence of protein production on the proteins at other vertices. This work continues with the maps of Kohn[12], containing important information for study and deserving citation here.

The case of dynamical systems theory is discussed in a paper by Mendes[13]. Here the dynamical behavior of "agents" placed in the nodes with interactions defined by the link structure of the networks. The node dynamics are modeled by ordinary differential equations much like chemical kinetics is usually expressed. An illustration of this is shown for the p53 gene acting through a complex network of interactions whose description is simplified. The importance of the path that the network takes to reach its state is emphasized, rather than just the structure of the network.

Also of importance is the work of fractal genomics[14], which describes a new way to look into the dynamics of complex networks. It does take into account the long-range order representing self-similar or fractal behavior across different time scales and can derive time-based data of biological systems leading to scale-free network structure. It can represent which genes can "drive" their interactions.

## IX.  Property Wavelets

The preceding graphs show how the properties vary with the sequence of amino acid residues in the sequence, albeit it was only extended to nearest neighbors in the sequence. It is noted that there is a rapid *glissando* up and down the "scale" of amino acid residues along the protein. It could be likened to a musical score if we show the trends of the points or "notes" on the graph. Further more if we concentrate on the properties of the protein rather than its structure we can consider this property wavelet rather than the conventional representation. Such an optimized protein may have a very sort wave length and high frequency especially if it is to fit inside of a cell (likening it to the "particle in a box" problem of quantum mechanics where the nodes of the wavelet must be at the boundary of the cell. Indeed when we use the program Knapsack for a length of 10 residues as in the above example we find when we optimize for a property (such as log of the partition coefficient or hydrophobicity at minimum volume) that the result is very close to that of the human (Man) in the above example. The discrepancy may be a fault of errors in the data.

After this paper was written, the paper by de Trad et alia[15] was seen where CWT (Continuous Wavelet Transform) was used to analyze the informational content of proteins. These functions were used to compare their functions. It expanded conventional sequence similarity that only accounted for local pairwise amino acid match and ignored courser spatial resolutions. Again these studies involved structures and not properties.

Another most interesting paper that was read after this paper was written was that of Frappat et alia[16], which denoted the probability, *f(n)*, of finding a particular codon, XZN, in the *n*-th position as a function of rank. Here the rank of the probability distribution is deemed as some function of the energy and that the system satisfies periodic conditions and that the interaction with the environment can be modeled as the switching on of a constant magnetic field. Then the energetic levels are quantified and appear as a wave function as of a one dimensional harmonic oscillator. A toy model is visualized and it is surmised that there is an indication that a simple mathematical scenario, able to reproduce the observed distribution, provides us with an indication of strong physico-chemical constraints that add to the random effects. Indeed, the approach in this paper hopes to yield substantial substance to the frequency of an amino acid occurring in fixed positions. The data then that is shown in their paper also adds to the propriety of this paper. Also, it is suggested that the plateau shown in the plot of table 8 shown in their paper may be an order-disorder[17] critical region of the rank distributions.

## X. The Darwin-Fowler Method[18,19] and Conformal Mapping[20]

Suppose we have a system with partitions $a_0, a_1, a_2 \ldots a_r$ each with energy $E_0, E_1, E_2 \ldots E_r$ and weight $g_0, g_1, g_2 \ldots g_r$. The total number of permutations is $v!$ which is equal to $La_0! \, a_1! \, \ldots \, a_r!$ Then $L = v! / (a_0! \, a_1! \, \ldots \, a_r!)$. The weight of each state is the same and has the value $w = (v! \, g_0^{a_0} \, g_1^{a_1} \, \ldots \, g_r^{a_r}) / (a_0! \, a_1! \, \ldots \, a_r!)$. The value of $a_r$ and the energy $E_r$, can then be found by a method called the Darwin Fowler method, which arises from a branch of mathematics called "The method of steepest descents" and employs the finding the residue in complex variable theory.

Complex variable mathematics also is capable of conformal mapping. The Schwarz-Christoffel transformation[20] maps a polygonal boundary of the *z*-plane ($z = x + iy$) on the *t*-plane. This is done using the differential equation $dz/dt = A(t-r_1)^{-\alpha_1}(t-r_2)^{-\alpha_2}\ldots$ or $dz/dt = A\prod_{k=1}^{n}(t-r_k)^{-\alpha_k}$ where the $\alpha_k$ are fixed points on the *t*-plane and $\alpha_k\pi$ is the angle change during the transit around $r_k$. Now if the folding of a protein can be determined by just *n* values (*n* = 3 for 3 sequential amino acid residues in the protein chain, then the configuration of the fold of the protein polymer can be assigned to one of many known protein native states. Also if the ligand's position (that is adsorbed (docked) by the target protein or DNA) with respect to the target than the Darwin-Fowler method can be used to map conformably the target-ligand (or vertex-edge) conjunction into a specific network of proteins or protein-DNA network.

## XI. Conclusions

Although the literature emphasizes hydrophobicity as primarily important in the folding of proteins, neglecting other physical properties, this work includes other properties for determining the sequence of amino acid residues and finds them of value. When each of the physical properties of the amino acid residues are ranked according to relative values, the sequence is found to fit a function of physical properties including dipole moment, log of the partition function, log of the activity coefficient and solubility parameter averaged as a class, and the molecular volume and hydrophobicity as two other classes. The latter two are also considered as separate classes but the log of the partition coefficient is very closely allied with the hydrophobicity. Biological networks and evolutionary considerations are also important in the complete proteomic description. PNS and P-group techniques can help to delineate biological networks and the evolutions of protein structure of various organisms are traced by their physical properties and sequence structure.

Acknowledgements: Professors L. T. Fan of the Chemical Engineering Department of Kansas State University and Ferenc Friedler of the Computer Science Department of Veszprem University, Hungary, have been a great inspiration to me and I thank Laszlo Papp for all his computer acumen.

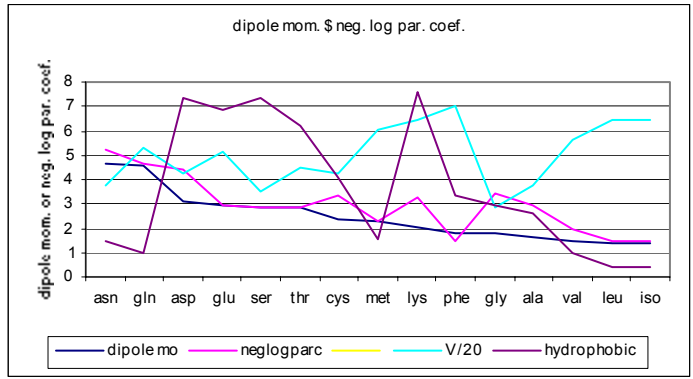

**Figure 1**

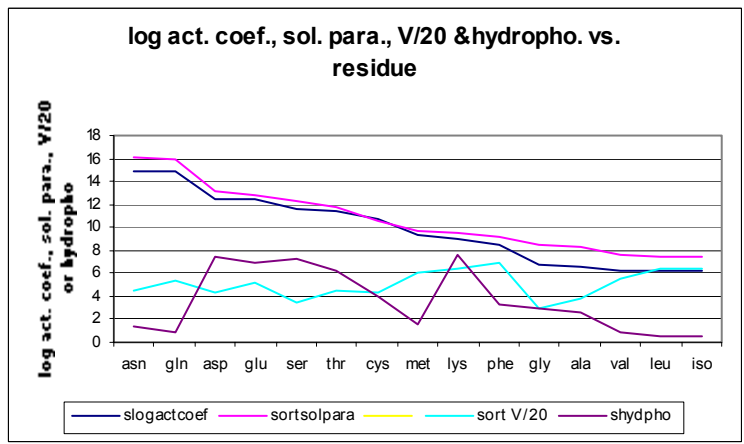

**Figure 2**

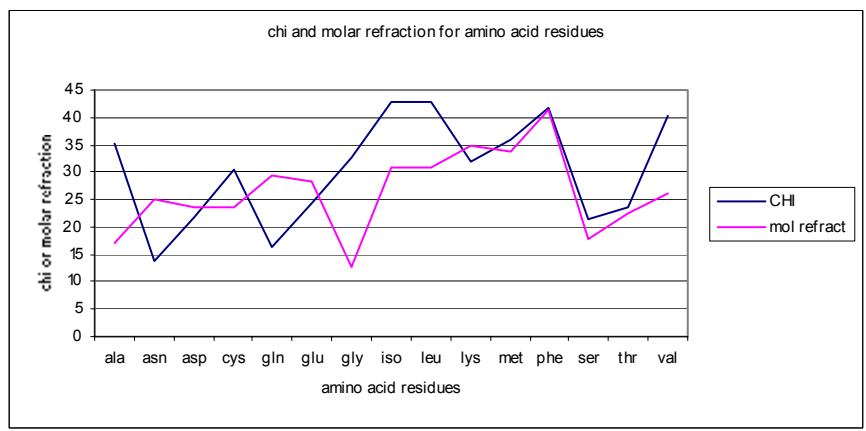

**Figure 3**

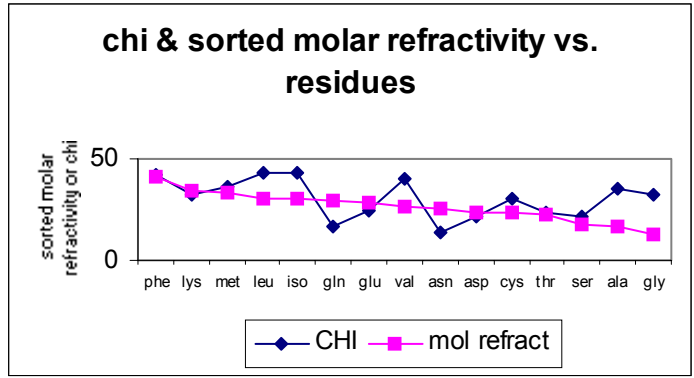

**Figure 4**

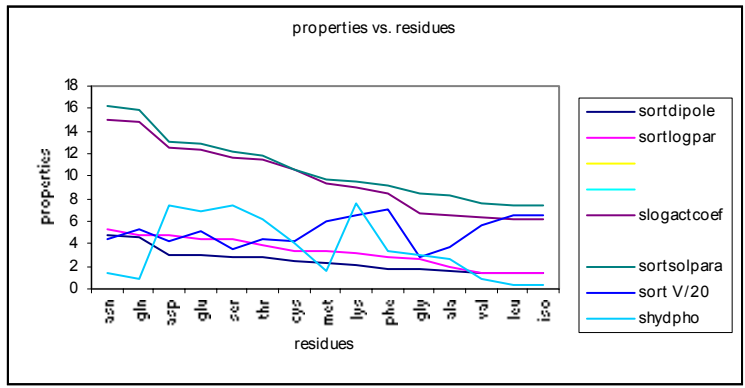

**Figure 5**

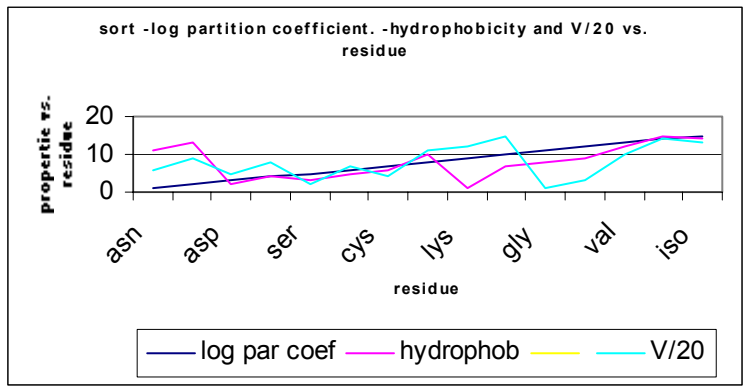

**Figure 6**

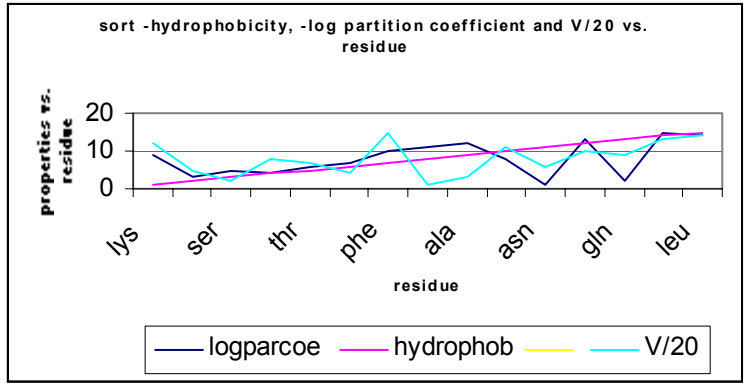

**Figure 7**

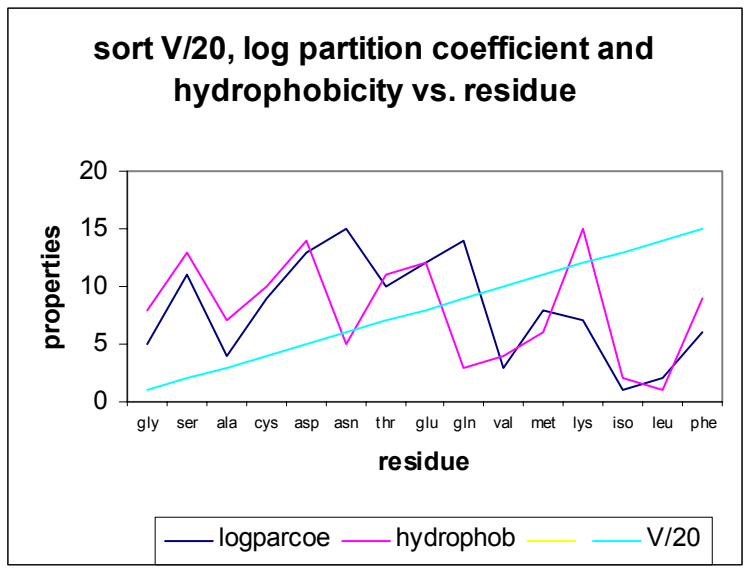

**Figure 8**

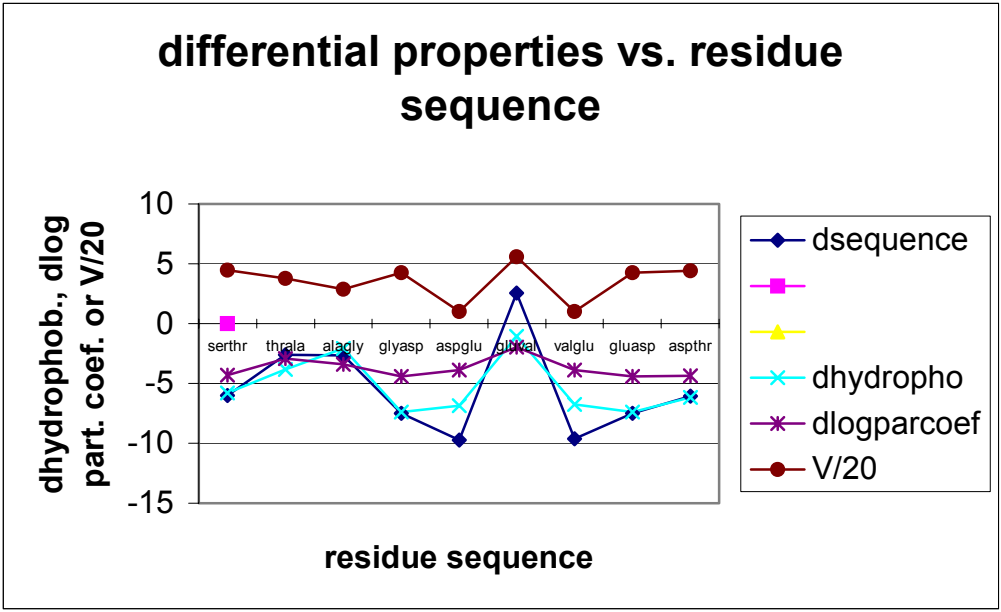

**Figure 9**

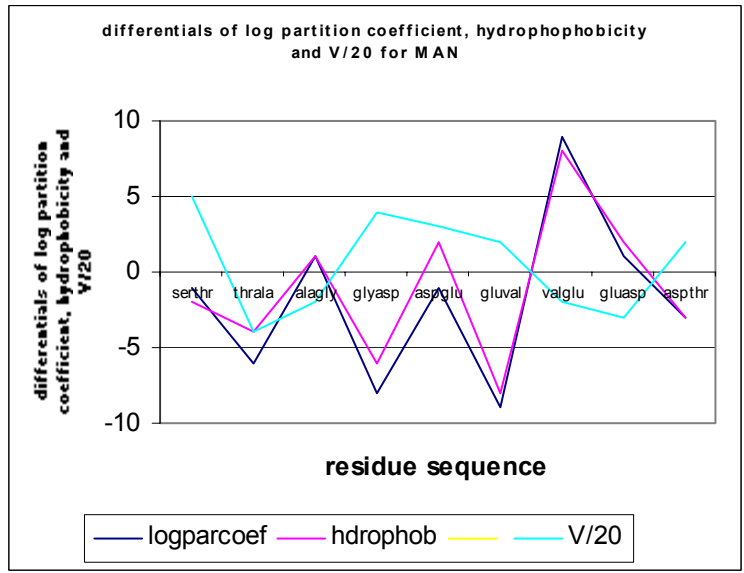

**Figure 10**

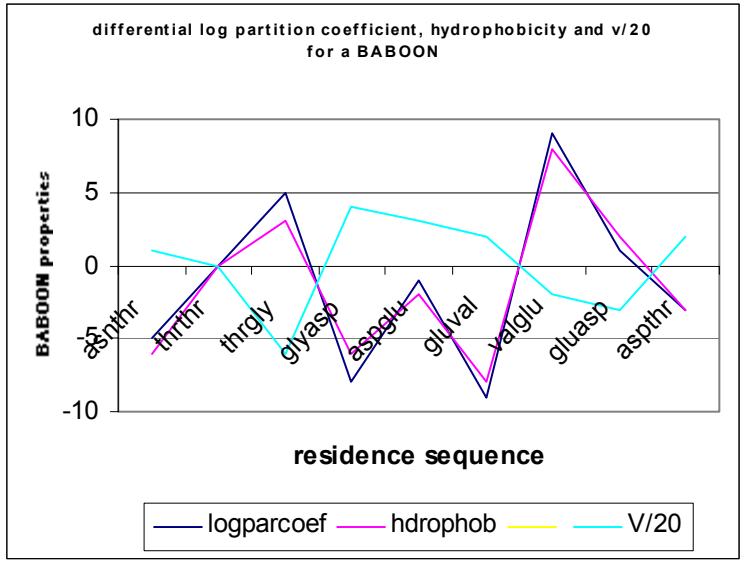

**Figure 11**

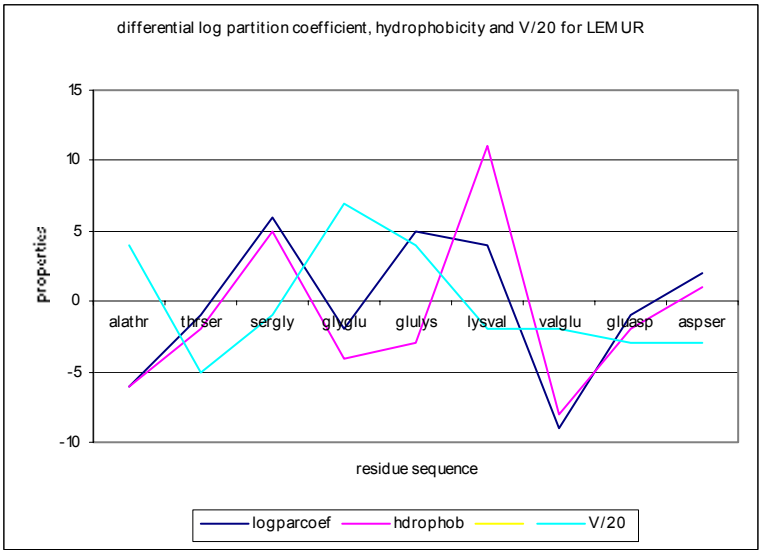

**Figure 12**

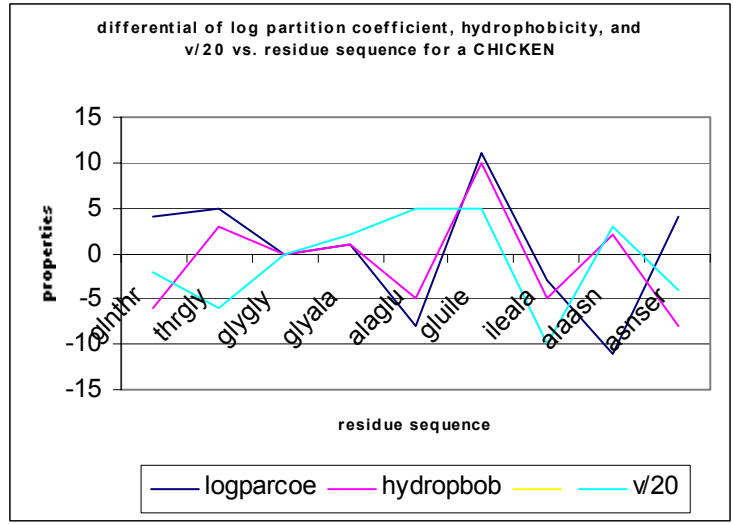

**Figure 13**

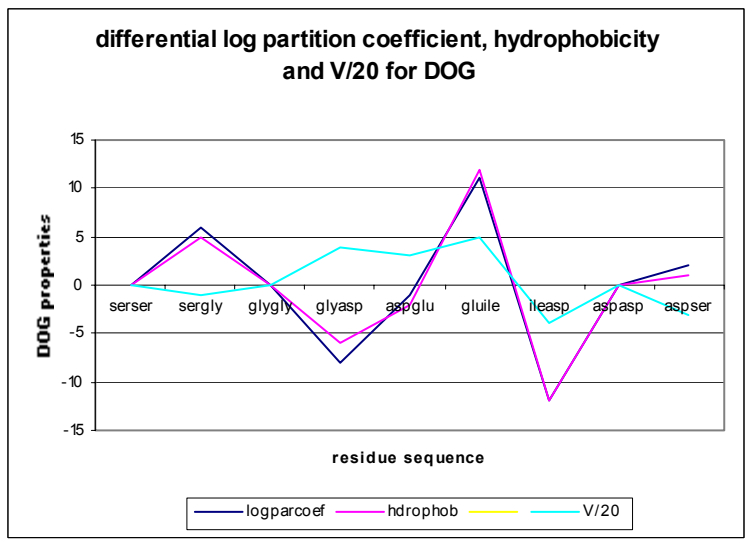

**Figure 14**